# Observation of quantum temporal correlations well beyond Lüders bound

Chun-Wang Wu,[1,2] Man-Chao Zhang,[1,2] Yan-Li Zhou,[1,2] Ting Chen,[1,2] Ran Huang,[3] Yi Xie,[1,2] Wen-bo Su,[1,2] Bao-Quan Ou,[1,2] Wei Wu,[1,2,4] Adam Miranowicz,[3,5] Franco Nori,[3] Jie Zhang,[1,2,*] Hui Jing,[6,†] and Ping-Xing Chen[1,2,4,‡]

[1]*Institute for Quantum Science and Technology, College of Science, NUDT, Changsha 410073, Hunan, China*
[2]*Hunan Key Laboratory of Quantum Information Mechanism and Technology, NUDT, Changsha 410073, Hunan, China*
[3]*Quantum Computing Center and Cluster for Pioneering Research, RIKEN, Wako-shi, Saitama 351-0198, Japan*
[4]*Hefei National Laboratory, Hefei 230088, Anhui, China*
[5]*Institute of Spintronics and Quantum Information, Faculty of Physics, Adam Mickiewicz University, 61-614 Poznań, Poland*
[6]*Department of Physics and Synergetic Innovation Center for Quantum Effects and Applications, Hunan Normal University, Changsha 410073, China*

The ability of achieving strong quantum spatial correlations has helped the emergence of quantum information science. In contrast, how to achieve strong quantum temporal correlations (QTCs) has remained as a long-standing challenge, thus hindering their applications in time-domain quantum control. Here we experimentally demonstrate that by using a parity-time ($\mathcal{PT}$)–symmetric single ion, the conventional QTC limit known as the Lüders bound can be well surpassed within a standard measurement scenario, approaching the predicted maximum QTC value. Our work, as a step toward quantum engineering of $\mathcal{PT}$ devices in the time domain, can stimulate more efforts on operating various quantum devices with the aid of strong QTC resources.

Quantum correlations are cornerstones in quantum information science and technology [1,2]. Among them, correlations between spatially separated systems can be benchmarked by the famous Bell inequalities [3], while quantum temporal correlations (QTCs) between temporally separated states of a system can be benchmarked by the Leggett-Garg inequalities [4]. Due to their exceptional ability to distinguish between quantum and classical phenomena, QTCs have proven to be useful in discriminating quantum transport [5,6], witnessing quantum non-Markovianity [7,8], and providing resources for quantum processing [9]. In quantum information science, the acceleration dynamics and parameter sensitivity behind strong QTCs can be used to achieve fast quantum state transfer [10] and enhanced quantum sensing [11]. QTCs can also be used for testing the security of practical quantum key distribution in quantum cryptography [12]. As an experimentally feasible indicator of quantum superpositions, QTCs even have the potential to test the long-lived quantum coherence in photosynthetic light harvesting [13–15]. In principle, an upper limit of QTCs was predicted but till now has remained a challenge to reach in practice; instead, only half of that limit, known as the Lüders bound, has been achieved within the standard measurement scenario [16–25], hindering deep understanding and more applications of quantum effects in the time domain.

Here we report an experimental observation of strong QTCs near the predicted maximum value, by using a parity-time ($\mathcal{PT}$)–symmetric qubit. We show that for such a nonlinear system, accelerated dynamics can induce enhanced QTCs well beyond the Lüders bound (Fig. 1). In particular, we observe the emergence of an exceptional point (EP) in our system and find that by approaching it, the maximum QTC value can be realized within the standard measurement scenario. We note that in previous works, quantum EP effects have been studied in a wide range of systems [11,26] but still merely in the time domain. Our work fills in this obvious gap and can stimulate more efforts on quantum engineering of EP devices in the time domain. A direct application example is the use of extreme acceleration dynamics that support strong QTCs in $\mathcal{PT}$-symmetric systems to speed up the generation of quantum entanglement [27].

QTCs can be quantified by the violation extent of the temporal Bell inequality, which Leggett and Garg proposed [4] for the original purpose of detecting macroscopic coherence. For a dichotomic observable $Q = \pm 1$, the temporal Bell inequality reads

$$K_3 = C_{12} + C_{23} - C_{13} \leqslant 1, \qquad (1)$$

where $C_{ij} = \langle Q(t_i)Q(t_j) \rangle$ is the two-time correlation function at times $t_i$ and $t_j$. The allowed ranges of the Leggett-Garg parameter $K_3$ for classical, Hermitian, and non-Hermitian quantum systems are shown in Fig. 1(a). It should be noted that, in the Hermitian quantum case, $K_3$ beyond the Lüders bound can be observed by devising delicate modifications to the standard Leggett-Garg test scenario, such as introducing multiple-time correlation functions [28], or using multiple-projector measurements [29–33]. Instead, here we focus on the framework of the standard measurement scenario.

---

*Contact author: zj1589233@126.com
†Contact author: jinghui@hunnu.edu.cn
‡Contact author: pxchen@nudt.edu.cn

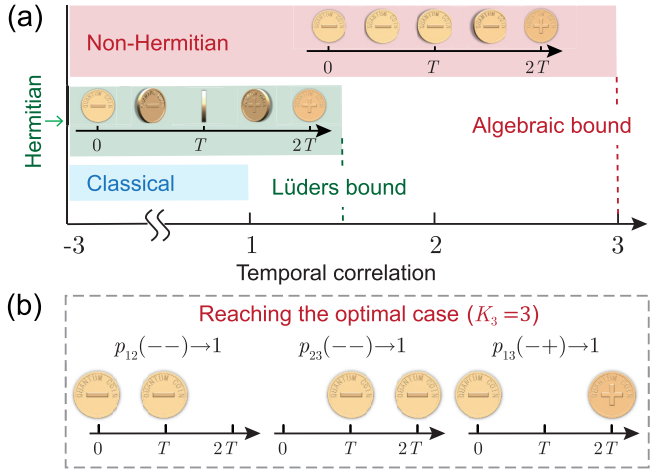

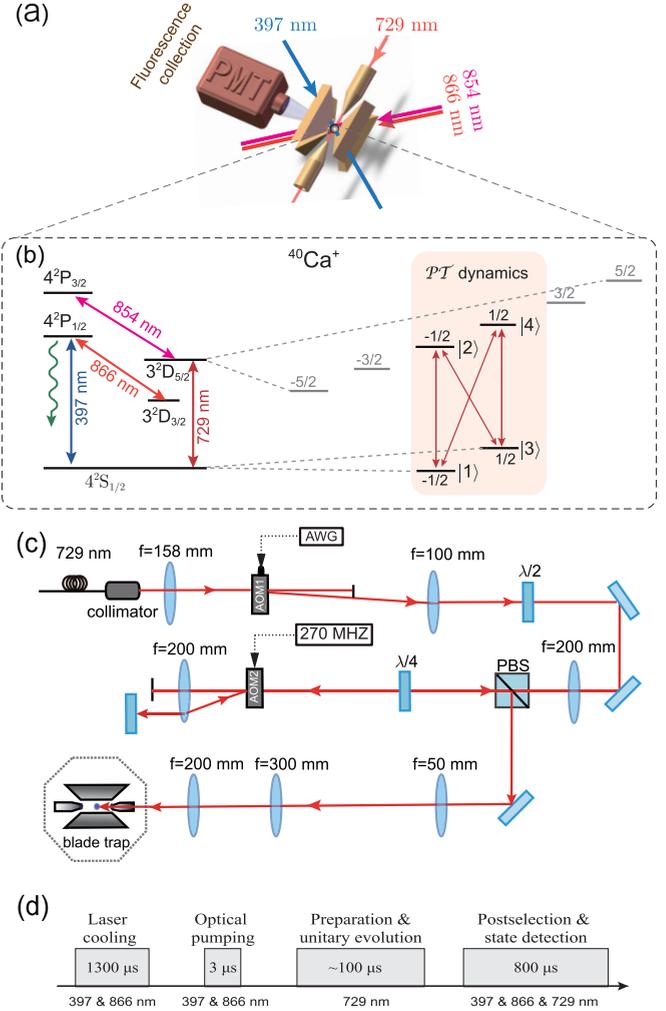

FIG. 1. Allowed ranges of QTCs for different types of physical systems, and the enhancement mechanism of temporal correlations by non-Hermiticity. (a) QTCs for classical, Hermitian quantum, and non-Hermitian quantum systems are bounded by 1, the Lüders bound ($K_3 = 1.5$), and the algebraic bound ($K_3 = 3$), respectively. (b) Successive measurements performed in time pairs on the non-Hermitian flip dynamics give a counterintuitive combination of the joint probabilities for measurement outcomes, resulting in the algebraic bound of QTCs, $K_3 = 3$.

As shown in Figs. 1(a) and 1(b), the mechanism of enhanced QTCs can be illustrated by comparing the Hermitian and non-Hermitian systems [34–36]. For a quantum coin, it can be in the superposition state $|\psi\rangle = \cos(\frac{\delta}{2})|-\rangle + \sin(\frac{\delta}{2})|+\rangle$, and the measurement makes it collapse to $|-\rangle$ and $|+\rangle$ with probabilities $p_- = \cos^2(\frac{\delta}{2})$ and $p_+ = \sin^2(\frac{\delta}{2})$, obtaining the observed values $Q = \pm 1$. In the Hermitian case, without measurement, the coin flips coherently from $|-\rangle$ to $|+\rangle$ with a uniform speed in a time period $[0, 2T]$, where $T$ is half the time required for the entire flipping process. However, in the non-Hermitian case, the coin can finish the same flip with a highly nonuniform speed, which is infinitesimal in most of the times $t < 2T$, but infinitely large in the neighborhood of $2T$. A measurement performed at the time $t < 2T$ can destroy the coherence and collapse the state to $|-\rangle$ with a probability close to 1, leading to a reset of the flip to the start. To implement the Leggett-Garg test, we choose $t_1 = 0$, $t_2 = T$, and $t_3 = 2T$, and define $p_{ij}(Q_i, Q_j)$ as the joint probability for the outcomes $Q_i$ and $Q_j$ of quantum measurements performed at times $t_i$ and $t_j$, respectively. Then, successive measurements performed in pairs give $p_{12}(-,-) \sim 1$, $p_{23}(-,-) \sim 1$, and $p_{13}(-,+) \sim 1$, resulting in the maximal QTC corresponding to $K_3 = 3$. In our experiment, $|+\rangle$ and $|-\rangle$ correspond to different superpositions of two Zeeman levels of a $^{40}$Ca$^+$ ion. The projection measurements on this basis can be realized using fluorescence detection assisted by quantum rotations, and the desired non-Hermitian dynamics can be embedded into a larger Hermitian system using the dilation method [37]. The joint probabilities $p_{ij}(Q_i, Q_j)$ are obtained in the prepare-and-measure scenario [6]. Then by approaching the exceptional point (EP) of an engineered $\mathcal{PT}$-symmetric Hamiltonian (a special class of non-Hermitian Hamiltonians), we can experimentally realize the aforementioned "first extremely slow, then infinitely fast" flip process [Fig. 3(d)] and achieve the upper limit of QTCs [Fig. 4(d)].

FIG. 2. Experimental layout. (a) A single $^{40}$Ca$^+$ ion trapped in a linear Paul trap. (b) Relevant energy levels used in our experiment. (c) Optical setup for the 729 nm laser, which is modulated by two acousto-optical modulators (AOMs) with a single-pass and a double-pass configuration, respectively. (d) Experimental procedure.

Our system can be described in the simplest level by an effective $\mathcal{PT}$-symmetric Hamiltonian

$$\hat{H}_{\text{PT}} = J\hat{\sigma}_x + i\Gamma\hat{\sigma}_z, \quad (2)$$

where $\hat{\sigma}_x = |1\rangle\langle 2| + |2\rangle\langle 1|$ and $\hat{\sigma}_z = |1\rangle\langle 1| - |2\rangle\langle 2|$ are Pauli operators, $J$ is the coupling rate of the two energy levels, and $\Gamma$ denotes the rate of the balanced gain (on $|1\rangle$) and loss (on $|2\rangle$). Clearly, $[\hat{H}_{\text{PT}}, \hat{\mathcal{P}}\hat{\mathcal{T}}] = 0$. For $J > \Gamma$, we have the $\mathcal{PT}$-symmetric (PTS) region and, similarly to the Hermitian case, the populations of $|1\rangle$ and $|2\rangle$, denoted

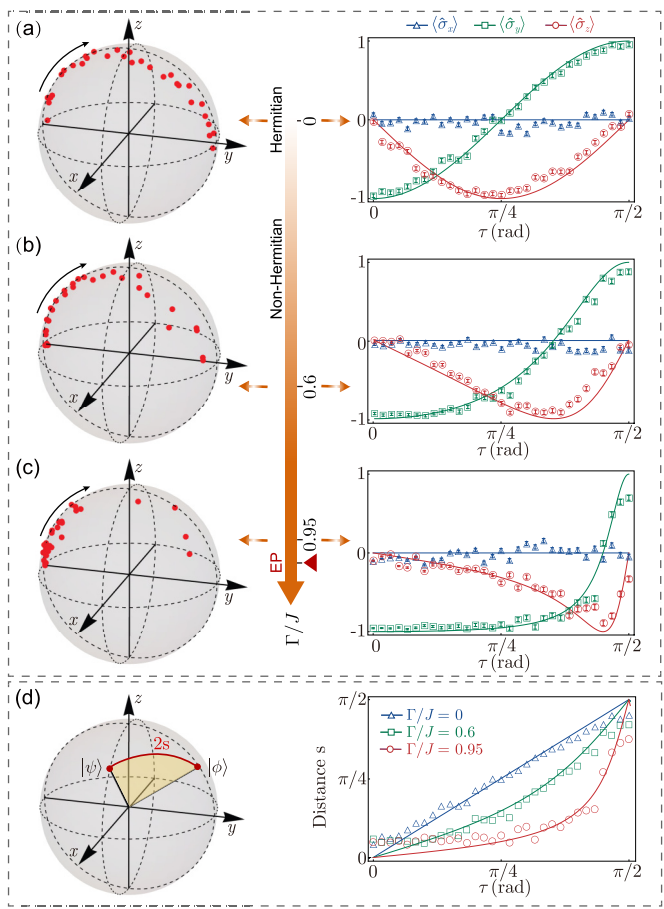

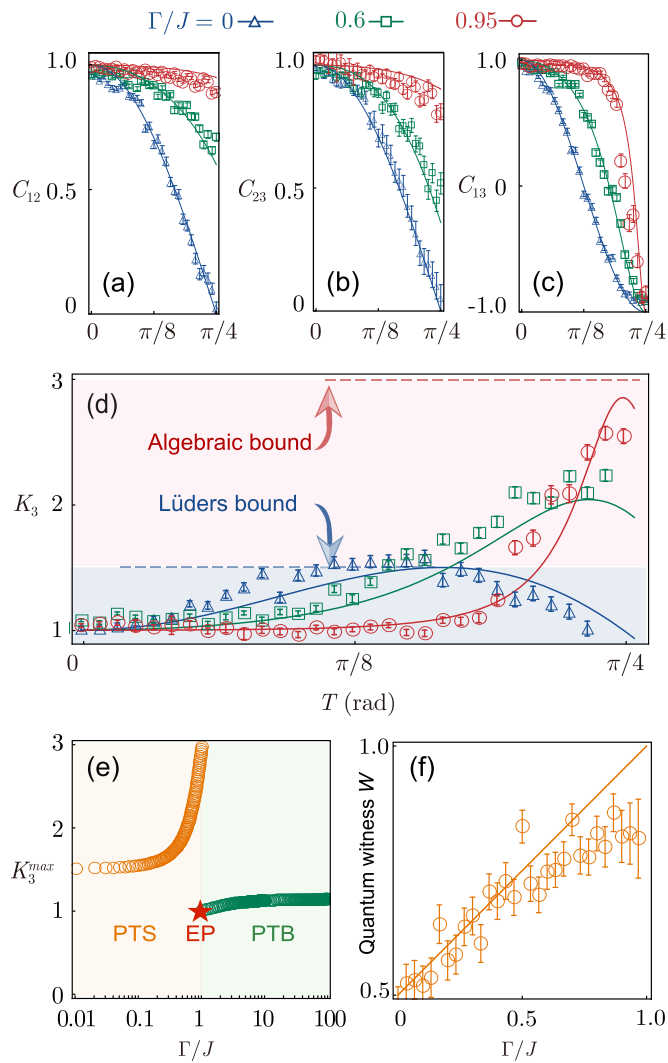

FIG. 3. Accelerated dynamics manifested by the $\mathcal{PT}$-symmetric qubit. (a)–(c) The Bloch vector (left) and its components (right) versus time $\tau$ for $\Gamma/J = 0$ (a), $\Gamma/J = 0.6$ (b), and $\Gamma/J = 0.95$ (c), when driving the qubit from $|-\rangle_y$ to $|+\rangle_y$. (d) The distance $s$ from the initial state, defined as half the geodesic distance on the Bloch sphere, is experimentally measured versus $\tau$ for various $\Gamma/J$, which shows a nonuniform flip speed in the non-Hermitian realm. The success rate of the postselection decreases with the increasing ratio $\Gamma/J$, and gradually vanishes when approaching EP. For $\Gamma/J = 0.95$ and $\tau = \pi/2$, the probability of successful selection is as low as 2.5%.

FIG. 4. Leggett-Garg parameter $K_3$ and quantum witness $W$ for the $\mathcal{PT}$-symmetric qubit. (a)–(d) The measured correlation functions $C_{12}$ (a), $C_{23}$ (b), $C_{13}$ (c), and the Leggett-Garg parameter $K_3$ (d) are plotted versus the measurement time interval $T$ for $\Gamma/J = 0$, 0.6, and 0.95. This shows that $K_3$ asymptotically approaches its algebraic bound of 3 when close to the EP. (e) With the fixed initial state $|-\rangle_y$ and the measurement observable $\hat{\sigma}_y$, the maximal QTC obtained by our theoretical optimization of the time interval is plotted in both the $\mathcal{PT}$-symmetric unbroken (PTS) and broken (PTB) regimes. (f) The experimentally measured quantum witness $W$ for various values of $\Gamma/J$, which also asymptotically approaches its algebraic bound of 1 when close to the EP.

as $p_1$ and $p_2$, exhibit Rabi-like oscillations with frequency $\Omega = \sqrt{J^2 - \Gamma^2}$. However, unlike the Hermitian case, the total population $p = p_1 + p_2$ also exhibits periodic oscillations (see similar features in classical $\mathcal{PT}$-symmetric systems [38]). The normalized density operator $\hat{\rho}(t)$ of this system satisfies the nonlinear equation [39],

$$\dot{\hat{\rho}} = -iJ[\hat{\sigma}_x, \hat{\rho}] + \Gamma\{\hat{\sigma}_z, \hat{\rho}\} \underbrace{- 2\Gamma\hat{\rho}[\text{Tr}(\hat{\sigma}_z\hat{\rho})]}_{\text{nonlinear term}}. \quad (3)$$

Here the nonlinear term can lead to an uneven speed of flip up and down, a feature not existing in linear Hermitian dynamics.

As shown in Fig. 2, our experiment is performed with a $^{40}\text{Ca}^+$ ion in a blade-shaped Paul trap. All the relevant energy levels and laser wavelengths involved in this experiment are shown in Figs. 2(a) and 2(b). The Zeeman sublevels $S_{1/2}(m_J = -1/2)$, $D_{5/2}(m_J = -1/2)$, $S_{1/2}(m_J = 1/2)$, and $D_{5/2}(m_J = 1/2)$ in a magnetic field of 5.3 G are chosen as quantum states $|1\rangle$, $|2\rangle$, $|3\rangle$, and $|4\rangle$, respectively, and are used to simulate the active $\mathcal{PT}$-symmetric dynamics. The transitions between these sublevels have least sensitivity to the fluctuations of the magnetic field, which allows these energy levels to have a coherence time of several milliseconds. The coherent couplings between these states are realized by applying a narrow-linewidth 729 nm laser beam with its wave vector along the axial direction of the linear Paul trap. Although only carrier transitions are used in this experiment, we still use the Doppler cooling and electromagnetically induced transparency (EIT) cooling methods [40] to prepare the motional degree of the freedom to the ground state, which guarantees high-fidelity operations of the single $\mathcal{PT}$-symmetric qubit. The single-pass AOM (labeled AOM1) is driven by an arbitrary wave generator (AWG) to control the intensity, frequency, and phase of the laser pulses, while the double-pass AOM (labeled AOM2) is used to shift the laser frequency close to the resonance of the $S$-$D$ transition [Fig. 2(c)]. The 397 nm laser is used for Doppler cooling, EIT cooling, and fluorescence detection. The 866 nm and 854 nm lasers are used for pumping the ion out of the $D$ states. To reduce the influence of the ac Stark effect, the Rabi frequency of each transition is set to about $2\pi \times 10$ kHz, which induces an estimated ac Stark shift of less than 50 Hz in the experiment. In this case, we can ignore the influence from the ac Stark shift.

The equatorial rotations $R(\theta, \phi) = \exp\{-i\theta[\cos(\phi)\sigma_x^m + \sin(\phi)\sigma_y^m]/2\}$ on any $S$-$D$ transition can be realized by resonantly driving the corresponding transition line using a narrow-linewidth laser at 729 nm, where $\theta$ is the rotation angle, $\phi$ is the laser phase, while $\sigma_x^m$ and $\sigma_y^m$ are respectively the Pauli $x$ and $y$ matrices in the representation of $\hat{\sigma}_z$. Using the digital quantum simulation method, which has been widely used in experimental studies of $\mathcal{PT}$-symmetric dynamics [41–43], the state preparation and an arbitrary unitary operation can be decomposed into appropriate sequences of $S$-$D$ equatorial rotations. The states in the $S$ manifold can be identified via state-dependent fluorescence observed using a photomultiplier tube (PMT), while coupling the $S_{1/2}$ state to the short-lived state $P_{1/2}$ by a laser field at 397 nm. With the help of appropriate $S$-$D$ rotations, this photon fluorescence detection method enables us to access populations $p_1$ and $p_2$.

Using the dilation method [37], we embed the $\mathcal{PT}$-symmetric dynamics into a directed sum of two 2D Hilbert spaces $\mathcal{H}_S$ and $\mathcal{H}_A$, where $\mathcal{H}_S$ and $\mathcal{H}_A$ are expanded by $\{|1\rangle, |2\rangle\}$ and $\{|3\rangle, |4\rangle\}$, respectively. The effective $\mathcal{PT}$-symmetric Hamiltonian in our experiment is achieved through quantum coherent evolution of a multilevel trapped ion, combined with appropriate postselection, so this construction technique does not introduce any approximations. We first prepare the total system into the initial state $|\Psi(\tau = 0)\rangle = N(|\psi\rangle_0 \oplus \eta|\psi\rangle_0)$, where $\eta = \frac{1}{2}[J, -i\Gamma; i\Gamma, J]$ is the metric operator for $\hat{H}_{\text{PT}}$ satisfying $\eta = \eta^\dagger$ and $\eta\hat{H}_{\text{PT}} = \hat{H}_{\text{PT}}^\dagger\eta$, and $N$ is the normalization factor. The unitary operation

$$U = \begin{pmatrix} F & G \\ -G & F \end{pmatrix} \quad (4)$$

is then enforced on the composite space $\mathcal{H}_S \bigoplus \mathcal{H}_A$, where the block matrices are $F = \cos(\tau)I_2^m - i\frac{\Omega}{J}\sin(\tau)\sigma_x^m$ and $G = \frac{\Gamma}{J}\sin(\tau)\sigma_z^m$, and $\tau = \Omega t$ is the scaled time; $I_2^m$ is the 2D identity matrix, and $\sigma_z^m$ is the Pauli $z$ matrix in the representation of $\hat{\sigma}_z$. The parameter $\Gamma/J$ can be controlled by changing the angles of the equatorial rotations that are used to form $U$ (see Supplemental Material [44] for more details). Then a postselection procedure is performed to discard the cases where the space $\mathcal{H}_A$ is occupied, and we achieve the $\mathcal{PT}$-symmetric dynamics $\hat{U}_{\text{PT}}|\psi\rangle_0$ in $\mathcal{H}_S$.

The total experimental procedure can be summarized as shown in Fig. 2(d): After the Doppler cooling and EIT cooling, the motional ground state of the ion can be prepared. By using an optical pumping pulse for 3 μs, the electronic state of the ion can be initialized into $S_{1/2}(m_J = -1/2)$. Then, a sequence of equatorial rotations using a 729 nm laser is performed to prepare the target state and achieve $U$. Finally, assisted by appropriate quantum rotation operations, the postselection and state measurement can be performed using fluorescence detection.

Then as shown in Fig. 3, we use the quantum state tomography technique to reveal the flip dynamics of the qubit from $|-\rangle_y = \frac{1}{\sqrt{2}}(|1\rangle - i|2\rangle)$ to $|+\rangle_y = \frac{1}{\sqrt{2}}(|1\rangle + i|2\rangle)$. The time intervals between adjacent data points of the Bloch vector [left of Figs. 3(a)–3(c)] and its components [right of Figs. 3(a)–3(c)] are equal. The initial state and Hamiltonian confine the evolution to the $y$-$z$ plane of the Bloch sphere. Increasing $\Gamma/J$, the accelerated dynamics becomes visible, which can be reflected by the dense-to-sparse behavior of the Bloch vector data points. The Bloch vector components versus time, plotted for different values of $\Gamma/J$, show a good agreement with our theoretical predictions, and the asymmetric features in the curves of the $y$ and $z$ components when $\Gamma/J > 0$ also demonstrate the acceleration properties. By introducing the Fubini-Study metric $s = \arccos|\langle\psi|\phi\rangle|$ as the distance between quantum states $|\psi\rangle$ and $|\phi\rangle$ [45], we can indirectly study the characteristics of the evolution speed $v = \frac{ds}{d\tau}$ in the flip process [Fig. 3(d)]. For the qubit, the distance $s$ is geometrically half the geodesic distance between $|\psi\rangle$ and $|\phi\rangle$ on the Bloch sphere. It is shown that, when $\Gamma/J = 0$, $s$ increases linearly and the Hermitian qubit evolves with a uniform speed. In the non-Hermitian realm with $\Gamma/J > 0$, the slopes in the plots of $s$ show increasing trends of the evolution speed with increasing $\tau$. For $\Gamma/J \to 1$, i.e., close to the EP, a flip occurs, which is similar to the non-Hermitian quantum coin as shown in Fig. 1(a).

Inspired by the idea of the "prepare-and-measure scenario" in Refs. [6,25], the two-time correlation functions in our experiment are indirectly computed via the experimentally measured conditional probability, $p_\tau(Q'|Q)$, for observing a measurement outcome $Q'$ at a scaled time $\tau$ given that we deterministically initialize the qubit in the eigenstate $|Q\rangle$. For testing the temporal inequality, we prepare the qubit in $|-\rangle_y$, choose the three scaled time instants as $\tau_1 = 0$, $\tau_2 = T$, and $\tau_3 = 2T$, and set the Pauli operator $\hat{\sigma}_y$ as the physical observable; then the two-time correlation functions can be given by

$$\begin{aligned} C_{12} &= -p_\text{T}(+|-) + p_\text{T}(-|-), \\ C_{13} &= -p_{2\text{T}}(+|-) + p_{2\text{T}}(-|-), \\ C_{23} &= p_\text{T}(+|-)p_\text{T}(+|+) - p_\text{T}(+|-)p_\text{T}(-|+) \\ &\quad - p_\text{T}(-|-)p_\text{T}(+|-) + p_\text{T}(-|-)p_\text{T}(-|-). \end{aligned} \quad (5)$$

The experimental results for $C_{12}$, $C_{13}$, and $C_{23}$ as functions of the scaled time interval $T \in [0, \frac{\pi}{4}]$ for $\Gamma/J = 0$, 0.6, and 0.95 are given in Figs. 4(a)–4(c). Because the non-Hermiticity can cause the slow flip speed of the qubit at the beginning, a larger $\Gamma/J$ can make the qubit states at times $\tau_1$ and $\tau_2$ more positively correlated, and eventually leads to a larger value of the correlation function $C_{12}$ [Fig. 4(a)]. The measurement performed at time $\tau_2$ collapses the superposition state into a mixture of $|+\rangle_y$ and $|-\rangle_y$, making the subsequent evolution as slow as the initial time, so the plot of $C_{23}$ has a similar behavior to that of $C_{12}$ [Fig. 4(b)]. However, without performing such a measurement, the qubit always finishes a complete flip in the time period $[0, \frac{\pi}{2}]$ independently of $\Gamma/J$. Therefore, the plots $C_{13}$ converge to $-1$ when $T = \frac{\pi}{4}$ [Fig. 4(c)]. Based on these data of the two-time correlation functions, the Leggett-Garg parameter $K_3$ is plotted in Fig. 4(d). It is seen that, in the Hermitian regime ($\Gamma/J = 0$), $K_3$ is bounded by the Lüders bound of 1.5. However, in the non-Hermitian regime ($\Gamma/J > 0$), $K_3$ obviously breaks this upper bound. For $\Gamma/J = 0.95$, we experimentally observe the Leggett-Garg parameter as large as $K_3 = 2.57 \pm 0.08$.

Enhanced QTCs result from the accelerated dynamics in non-Hermitian systems. This type of dynamics and extreme QTCs can be observed not only near the EP in the PTS phase here, but also in the $\mathcal{PT}$-broken (PTB) phase. In theory, it has been proven that by optimizing QTCs in the complete parameter space related to the initial state, the measurement observable, and the time interval, the maximum $K_3$ monotonically increases to 3 at EP and always saturates this bound in the PTB phase [36]. However, to explore the region past EP using another dilation method [46], we need to simultaneously drive the $S$-$S$, $D$-$D$, and $S$-$D$ transitions of the $^{40}\text{Ca}^+$ ion, which is still a great challenge for the current trapped ion experiments. This work focuses on approaching the algebraic bound of QTC from the $\mathcal{PT}$-symmetric side. With this goal in mind, the fixed measurement observable $\hat{\sigma}_y$ and the initial state $|-\rangle_y$ have not only been proven to be a wise choice in theory [34–36], but also suitable to conduct experiments. Under this scenario, we demonstrated an interesting phenomenon shown in Fig. 4(e): with the same measurement configuration (the fixed initial state and measurement observable), the maximum $K_3$ obtained by our theoretical optimization of the time interval can exhibit discontinuity at the EP, which provides a source of inspiration for studying $\mathcal{PT}$-symmetric phase transitions from the perspective of QTCs.

The quantum witness $W = |p'(Q) - p(Q)|$ [47], another indicator of QTCs, is also tested in our experiment, where $p'(Q)$ and $p(Q)$ are the probabilities for observing the measurement outcome $Q$ after the period of the system's evolution in the presence and absence of an earlier measurement performed on the initial state, respectively. In particular, $W \leqslant \frac{1}{2}$ holds for a Hermitian qubit. In the experiment, we initialize the qubit in $|\psi\rangle_0 = (-\sqrt{J-\Gamma}|+\rangle_y + \sqrt{J+\Gamma}|-\rangle_y)/\sqrt{2J}$, choose the observable $\hat{\sigma}_y$, set the scaled evolution time prior to the later measurement as $\frac{\pi}{4}$, and compare the probabilities of obtaining the outcome $Q = +1$ in the two measurement scenarios. Then, the quantum witness $W$ for different values of $\Gamma/J$ is plotted in Fig. 4(f). It is shown that, in the Hermitian realm ($\Gamma/J = 0$), we experimentally obtain the Hermitian upper bound $\frac{1}{2}$. More interestingly, when we get close to the EP ($\Gamma/J \to 1$), the quantum witness asymptotically approaches its algebraic bound of 1. Similarly to the Leggett-Garg parameter, with appropriate initial state preparation and measurement observable, the quantum witness always saturates the algebraic bound in the PTB phase.

Note that in conventional Hermitian cases, to achieve the algebraic bound of QTC, we need in principle either the system size or the number of measurements to approach infinity. However, in our Leggett-Garg test, the number of measurements was only 3 and the dimension of the larger system where our $\mathcal{PT}$-symmetric Hamiltonian resides was 4. From an operational perspective, we show that the postselected dynamics of a low-dimensional system can be used to emulate infinite-dimensional systems in certain aspects of QTCs [34]. We expect that this strategy, as well as a stronger QTC achieved in our experiment, can stimulate more works on temporal quantum effects and their applications [48,49].

It should be noted that the $\mathcal{PT}$-symmetric qubit has been experimentally realized in trapped ion systems using passive loss channels [50,51], and in nitrogen-vacancy centers using the dilation method [46]. In those works, with the choice of the loss or gain state as the initial state, the demonstrated nonuniform evolutions are limited to skewed Rabi oscillations, which are unable to enhance the QTCs. In principle, after choosing the appropriate initial state and measurement observable, our strategy can be extended to these other experimental systems, enabling opportunities of exploring and utilizing QTCs in various EP systems. Additionally, QTCs in the $\mathcal{PT}$-broken region can be easily studied using the passive $\mathcal{PT}$ technique. However, using the passive ion can be affected by the unwanted population backflow to the qubit manifold, as verified by a very recent experiment in which experimental fidelities are severely affected and the QTC obtained is only up to 1.703 [52]. Another recent experimental work [53], which uses a passive $^{171}\text{Yb}^+$ ion to study both the three-order and four-order Leggett-Garg parameters $K_3$ and $K_4$, also suffers severely from the same backflow effect, which poses challenges for accurate detection. In contrast, here we implemented active $\mathcal{PT}$ dynamics using multilevels of a trapped ion, enabling a QTC value up to 2.57. In view of the records of quantum operation fidelities [54] and coherence time [55], we expect that the trapped ion system can serve as a powerful tool for exploring more interesting issues such as multiqubit QTCs or QTCs at higher-order EPs.

This work was supported by the National Natural Science Foundation of China (Grants No. 12174448, No. 12174447, No. 12074433, No. 12004430, No. 11904402, No. 12204543, and No. 11935006), the Innovation Program for Quantum Science and Technology (Grant No. 2021ZD0301605), the Science and Technology Innovation Program of Hunan Province (Grants No. 2022RC1194, No. 2023JJ10052, and No. 2020RC4047), and the Natural Science Foundation of Hunan Province (Grant No. 2023JJ30626). A.M. was supported by the Polish National Science Centre (NCN) under Maestro Grant No. DEC-2019/34/A/ST2/00081.